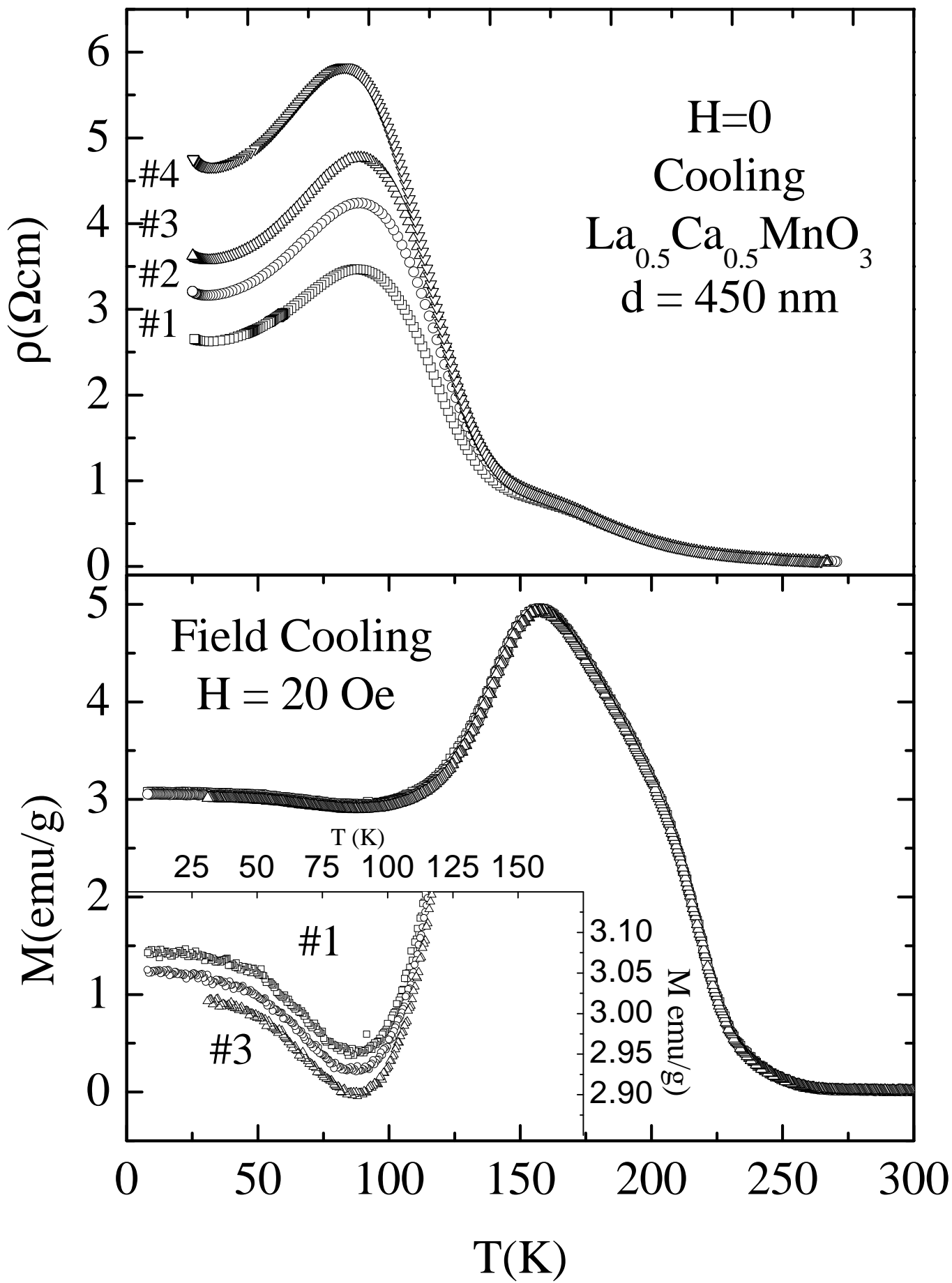

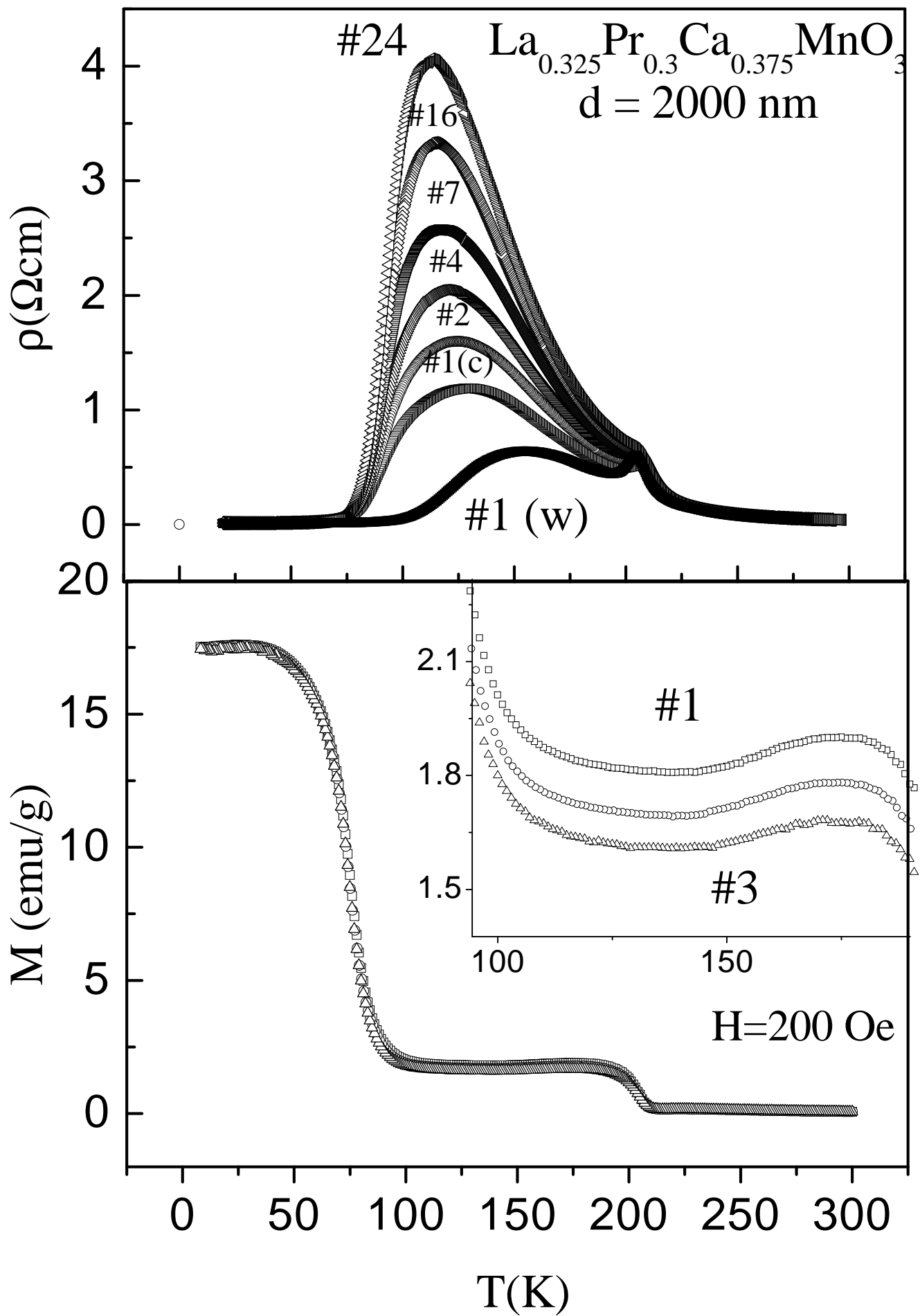

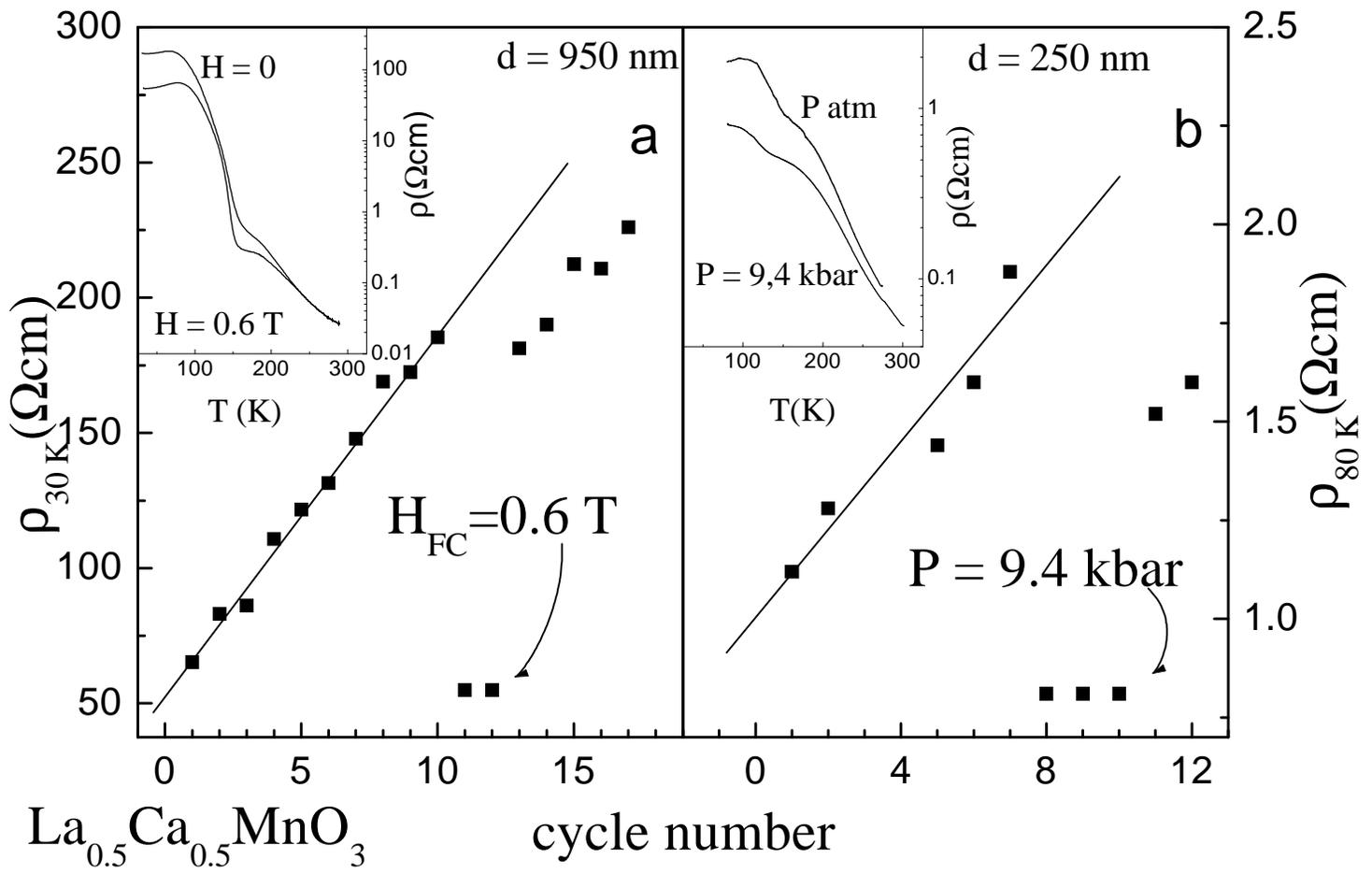

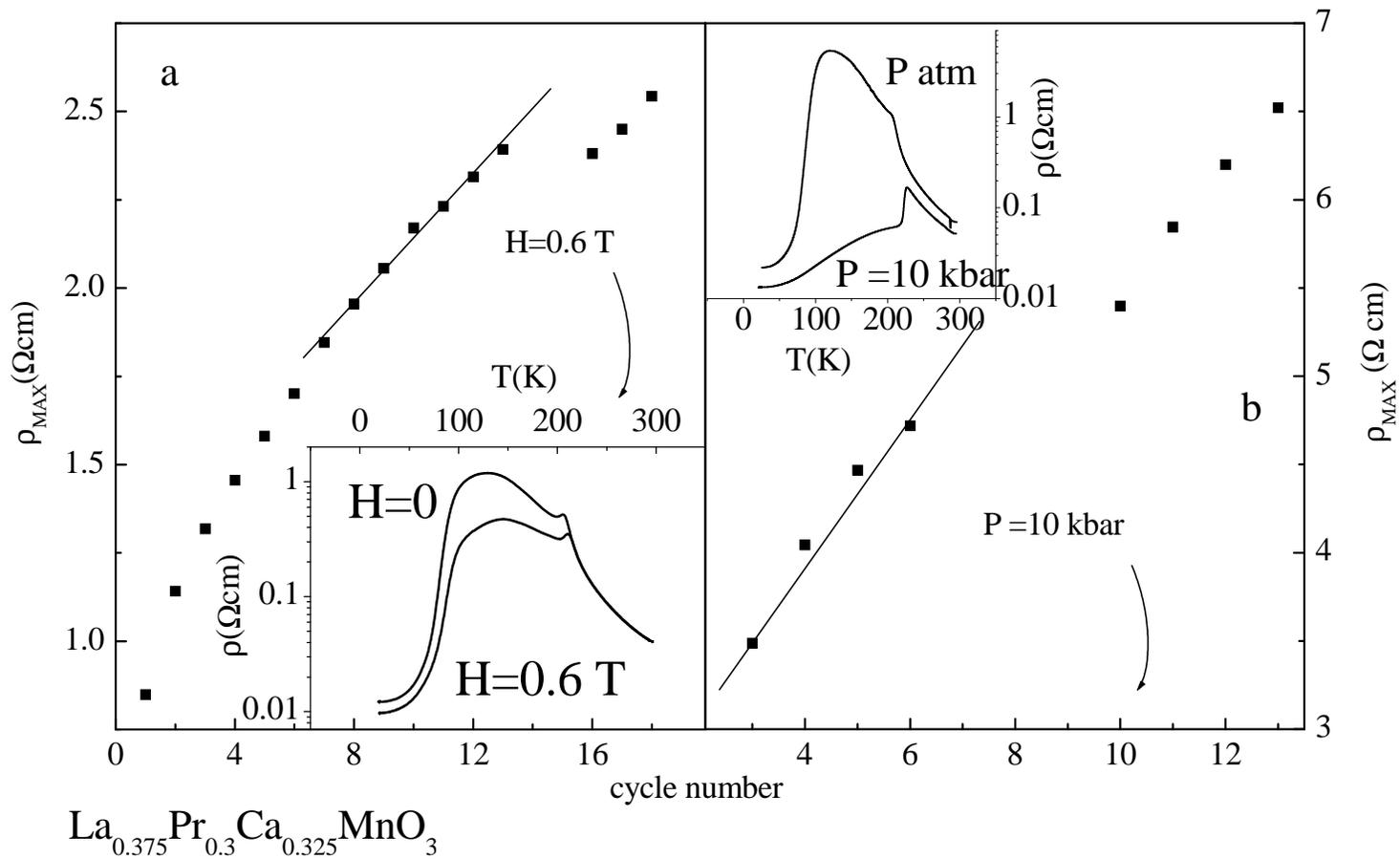

La$_{0.375}$Pr$_{0.3}$Ca$_{0.325}$MnO$_3$

# Low temperature irreversibility induced by thermal cycles on two prototypical phase separated manganites

J. Sacanell[a], M. Quintero[a], J. Curiale[a], G. Garbarino[b], C. Acha[b], R.S. Freitas[c], L. Ghivelder[c], G. Polla[a], G. Leyva[a], P. Levy[a], F. Parisi[a].

[a]Departamento de Física, CAC, CNEA, Av. Gral Paz 1499 (1650), San Martín, Buenos Aires, Argentina.
[b]Laboratorio de Bajas Temperaturas, Departamento de Física, FCEN, UBA, Argentina.
[c]Laboratorio de Baixas Temperaturas, Instituto de Fisica, UFRJ, Rio de Janeiro, Brasil.

We have studied the effect of irreversibility induced by repeated thermal cycles on the electric transport and magnetization of polycrystalline samples of $La_{0.5}Ca_{0.5}MnO_3$ and $La_{0.325}Pr_{0.3}Ca_{0.375}MnO_3$. An increase of the resistivity and a decrease of the magnetization at different temperature ranges after cycling is obtained in the temperature range between 300 K and 30 K. Both compounds are known to exhibit intrinsic submicrometric coexistence of phases and undergo a sequence of phase transitions related to structural changes.
Changes induced by thermal cycling can be partially inhibited by applying magnetic field and hydrostatic pressure. Our results suggest that the growth and coexistence of phases with different structures gives rise to microstructural tracks and strain accommodation, producing the observed irreversibility.
Irrespective of the actual ground state of each compound, the effect of thermal cycling is towards an increase of the amount of the insulating phase in both compounds.

Manganese oxides with perovskite structure and mixed valent Mn exhibit interesting features related to their electric, magnetic and structural properties, the colossal magnetoresistance effect is probably the best known of them.
The discovery of phase separation (PS) in some manganites [1], i.e. the intrinsic coexistence of two or more phases with different magnetic and transport properties on a submicrometric scale, provides an interesting scenario to account for some unusual properties. As thoroughly documented in the recent literature [1,2], this mixture is formed by a ferromagnetic (FM) conductive phase (zones in which double exchange coupling between $Mn^{+3}$ and $Mn^{+4}$ favours delocalization of carriers) and highly insulating charge ordered (CO) and even orbital ordered phases. Among manganites exhibiting macroscopic PS, $La_{5/8-y}Pr_yCa_{3/8}MnO_3$ [LPCMO(y)] with $0.3 < y < 0.4$, has its temperature of charge ordering ($T_{co}$) higher than the FM ordering one ($T_C$) while $La_{0.5}Ca_{0.5}MnO_3$ [LCMO] has $T_C$ higher than $T_{co}$.
On cooling LPCMO(0.3), nucleation of FM droplets below $T_C=200$ K occur in a CO host previously developed ($T_{co} = 230$ K). These clusters grow as the temperature is lowered, and an insulator-metal transition is obtained when the fraction of the FM phase reaches the percolation threshold. [2,3,4]
On the other hand, when cooling LCMO, CO regions nucleate within a charge delocalized host. This compound shows PS in the whole temperature range below $T_C \sim 220$ K. [5,6,7,8,9], with three well differentiated PS regimes [7].
In previous works we have observed magnetic memory effects in LPCMO(0.3) [10] and in B site doped LCMO [11] induced by external magnetic field H.
Recently, by measuring polycrystalline samples of $Pr_{0.5}Ca_{0.5}Mn_{1-x}M_xO_3$, (M = Cr, Co, Al, Ni) with x=0.03, Mahendiran et al. have shown that the low temperature resistivity increases after performing repeated thermal cycles.[12] In $La_{0.5}Ca_{0.5}MnO_3$, R-W. Li et al. have shown a training effect characterized by the reduction of the low temperature magnetization [13] and S. K. Hasanaian et al. have reported a similar effect by measuring ac susceptibility. [14] These authors agree that the effect is due to the appearance of strain and distortions localized at the interfaces between the FM and CO phases each time the sample is cycled.

In this work, we report the effect of repeated thermal cycles on two prototypical PS manganites [3] with rather different hole content, half doped LCMO and LPCMO(0.3). We have observed a persistent memory of the thermal history, namely changes in electric and magnetic properties at low temperature after each thermal cycle which remain after the sample is heated to room temperature. We have performed experiments with external H and hydrostatic pressure to get further insight on the effect.

Polycrystalline samples were synthesized by the sol gel technique, thermal treatments were performed at different temperatures (around 1000 C) to obtain pellets with different average grain size (d), which were determined through SEM micrographs. For comprehensive studying the effect of grain size on LCMO and LPCMO see ref. [6] and [4] respectively.
Resistivity (ρ) (standard four probe technique) and magnetization (M) (extraction QD PPMS magnetometer) were measured as a function of temperature T. Pressure up to 10 kbar was applied using a self-clamped cell.

Figure 1 shows several consecutive measurements of ρ and M for LCMO (d ≈ 450 nm) on cooling. As can be seen, after every thermal cycle between 300 and 30 K, the low temperature ρ increases approximately 1 Ωcm per cycle. Note that above 200 K, ρ is not modified by repeated cycling. When properly normalized, low T ρ data for different measurements on LCMO collapse into a single curve, suggesting that the same transport mechanism, but different amounts of coexisting phases are present in each cycle (see inset Fig. 1). Thus, this increase of ρ in the region where the compound is phase separated suggests that the fraction of coexisting phases changes as the sample is repeatedly cycled.

Concomitantly, M measured with a low H, decreases as thermal cycles are performed, unambiguously confirming a reduction in the amount of the FM fraction (*f*) after each one. If higher H (> 0.6T) are used to measure M, no reduction with cycles is observed.

It's worth noting that percentual changes in M are almost negligible as compared to those found in ρ. As close to the percolative threshold the effective ρ is mainly determined by the percolative character of *f* [2], "small" changes in *f* which may inhibit some key paths will produce huge changes in ρ but not on M, which is a volumetric measure of *f*.

As shown for LCMO in Fig. 1, a similar thermal cycling effect is found for LPCMO(0.3) (see Fig. 2) in the range 90 K < T < 200 K, notwithstanding the differences between the PS states of both compounds. Remarkably, only the phase coexistence region shows a change in measured properties, i.e., ρ below T ~ 90 K, where the compound is homogeneous FM, does not exhibit the anomalous increase in ρ after cycling. Also, the paramagnetic high T range (T > 230 K) is not affected, as happens for the LCMO sample. These results are in agreement with recently published data on PS compounds. [12,13,14]

The coexistence of two phases with different structures is expected to give rise to microstress localized at the interfaces as one phase is growing inside the other while a transition takes place [15], in close similarity with martensitic phase transitions. Following Podzorov et al. [4], the physics of the CO transformation is dominated by the accommodation strain induced by the nucleation of child regions within the parent matrix.

Thus, as PS is closely related to the very much similar energy scale for CO and charge delocalized FM interactions, small and local perturbations affect the relative fraction of coexisting phases.

Remarkably, the effect of thermal cycles is in both compounds towards a loss in the amount of the FM phase, irrespective of their actual ground state (PS for LCMO and FM for LPCMO(0.3)), suggesting that the training produced by crossing several times a structural transition introduces defects which stabilize non FM regions. Through this mechanism, the sample tracks its thermal history; this memory has an eminently structural character, displaying features similar to the shape memory effect found in other systems. [16]

We have further studied the effect of low H and hydrostatic pressure (P) on the increase of ρ produced by the thermal cycles, since both perturbations tend to stabilize FM regions.

For Mn – based oxides exhibiting PS, H aligns Mn magnetic moments, enhancing carrier itinerancy [1] and increases *f* [17]; P tends to inhibit Mn-O-Mn distortions which localize carriers [18].

When thermal cycles were performed either under applied H= 0.6 T or P= 9.4 kbar, the ρ increase effects were found to be negligible, as compared to those found without external perturbations. For comparison, Fig.3 shows the low temperature ρ of LCMO as a function of thermal cycle number, including the cycles in which H=0.6 T (Fig.3a) and P=9.4 kbar (Fig.3b) were applied. As shown, H inhibits the increase of the low T ρ while applied, but the effect resumes after subsequent thermal cycles without external stimulus (i.e. H=0). At this point, the increase of ρ starts from nearly the same value as the previous to the application of H one. Differently, the application of P=9.4 kbar on LCMO partially erases the effect of previously performed thermal cycles (Fig. 3b). This result enforces the microstructural origin of the observed instability.

Fig. 4 shows the effect of repeated thermal cycling on LPCMO(0.3), including cycles in which H= 0.6 T (Fig.4a) and P=10 kbar (Fig.4b) were applied. The general trend of results obtained is similar for both LCMO and LPCMO(0.3) compounds. The most noticeable difference appears after cycling under hydrostatic pressure, as for LPCMO(0.3) ρ is not reduced, but resumes increasing from a slightly higher value after the sample is cycled again under atmospheric pressure.

The surprising way in which P affects the thermal cycles effect suggest the microstructural origin of the observed phenomenon. Both systems undergo structural phase transitions that (from one cycle to another) tend to favour the CO state by reducing *f*. The role of the external P is to partially inhibit the effect of these phase transitions on LPCMO(0.3) and to reverse previous changes on LCMO by erasing the microstructural changes which appeared on previous cycles.

In summary, we have shown that the FM fraction decreases after performing a thermal cycle, irrespective of the ground state of the system under study, on two prototypical Mn – based oxides. The origin of this effect is most likely related to microstructural defects induced by the structural phase transition which provokes the growth of a certain phase in a matrix with different structural parameters. We have shown that the effect can be affected by both H and P, but after the removal of them, it is recovered on subsequent thermal cycles. The "quenching" effect of H is presumably due to the enlargement of $f$ [17] which overcomes the thermal cycle effect. The way in which P erases the thermal history of LCMO seems to support the idea of a structural change as the responsible for the thermal cycles effect. The difference between the effect of P on LCMO and LPCMO is a subject which is not clear at present and still more research is needed to better elucidate the structural driven uncontrolled change in the amount of phases through thermal cycles.

Figure 1: Consecutive measurements of ρ and M (measured with H = 20 Oe) as a function of temperature on cooling for LCMO (d ≈ 450 nm). After each cooling cycle, the sample was warmed up to room temperature. Labels indicate cycle number. Cooling rate was 2 K/min for ρ and 1 K/min for M. Inset: ρ data normalized to its low T value for consecutive cycles.

Figure 2: Consecutive measurements of ρ and M (measured with H = 200 Oe) as a function of temperature on cooling for LPCMO (grain size ≈ 2000 nm). After each cooling cycle, the sample was warmed up to room temperature. Labels indicate cycle number. Cooling rate was 2 K/min for ρ and 1 K/min for M.

Figure 3: a) ρ(30K) vs. cycle number, showing the effect of H on LCMO. Inset: ρ vs. T with H=0 and H=0.6 T on cooling. b) ρ(80K) vs. number of cycle after applying and relaxing P = 9.4 kbar on LCMO. Inset: ρ vs. T under P = 1 atm and P = 9.4 kbar on cooling.

Figure 4: a) $\rho_{MAX}$ vs. number of cycle, showing the effect of H on LPCMO. Inset: ρ vs. T under H=0 and H=0.6 T on cooling. b) $\rho_{MAX}$ vs. number of cycle after applying and relaxing P=10 kbar on LPCMO. Inset: ρ vs. T under atmospheric pressure and 10 kbar on cooling.